\documentclass[prl,preprint,epsfig,rotate,showpacs]{revtex4}
\usepackage{epsfig}
\usepackage{graphicx}
\usepackage{amsmath}

\newcommand{\beq}{\begin{eqnarray}}
\newcommand{\eeq}{\end{eqnarray}}
\begin{document}
\title{Comment on ``Quantification of Macroscopic Quantum Superpositions within Phase Space"} 
\author{Jiangbin Gong}
\affiliation{Department of Physics and Center for Computational
Science and Engineering, National University of Singapore, 117542,
Singapore}
\affiliation{NUS Graduate School for Integrative Sciences
and Engineering, Singapore 117597, Singapore}
\date{\today}
\maketitle

In a recent Letter~\cite{lee}, Lee and Jeong studied a phase space structure measure in order to
quantify macroscopic quantum superposition states. Their measure, claimed to be ``the most general and inclusive measure ever proposed",
was applied to different types of quantum states with interesting results.

Unfortunately, Lee and Jeong missed a direct connection between their measure (denoted ${\cal I}$) and a well-studied phase space measure (denoted $\chi^2$) in the literature~\cite{gu,brumer,gongpra,casati}.  In particular, the two measures are entirely equivalent for pure states and still closely related for mixed states.  

In Ref.~\cite{lee}, ${\cal I}$ is defined as
\begin{eqnarray}
{\cal I}\equiv \sum_{m=1}^{M}\text{Tr}
\left[\frac{1}{2}\rho^2\hat{a}^{\dagger}_m\hat{a}_m+\frac{1}{2}\rho\hat{a}_m^{\dagger}\hat{a}_m\rho-\rho\hat{a}_m\rho\hat{a}_m^{\dagger}\right],
\end{eqnarray}
where $\rho$ is the density operator of an $M$-mode system, $\hat{a}^{\dagger}$ and $\hat{a}_m$ are the creation and annihilation operators. Adopting dimensionless canonical pairs $(\hat{q}_m, \hat{p}_m)$ with $[\hat{q}_m,\hat{p}_{m'}]=i\delta_{mm'}$, we have $\hat{a}^{\dagger}=[\hat{q}_m-i\hat{p}_{m}]/\sqrt{2}$  and $\hat{a}_m=[\hat{q}_m+i\hat{p}_{m}]/\sqrt{2}$ and hence
\begin{eqnarray}
{\cal I}= \frac{1}{2}(C - M P),
\label{i1}
\end{eqnarray}
where
\begin{eqnarray}
C\equiv \sum_{m=1}^{M}\text{Tr}\left[ \rho^2 \hat{q}_m^2+ \rho^2 \hat{p}_m^2 -\rho\hat{q}_m\rho\hat{q}_m- \rho\hat{p}_m\rho\hat{p}_m\right],
\end{eqnarray}
and $P\equiv \text{Tr}[\rho^2]$ is the state purity.
In terms of the Wigner function $W\equiv [1/(2\pi)^{M}]\int d\mathbf{\eta}\langle {\bf q}+\mathbf{\eta}/2|\rho|{\bf q}-\mathbf {\eta}/2\rangle\exp(-i\mathbf{\eta}\cdot {\bf p})$ (${\bf q}$, ${\bf p}$ and ${\bf \eta}$ are $M$-dimensional vectors), we have
\begin{eqnarray}
C&=&\sum_{m=1}^{M} \frac{(2\pi)^{M}}{2} \int \left(\left|\frac{\partial W}{\partial q_m}\right|^2 + \left|\frac{\partial W}{\partial p_m}\right|^2\right)\ d{\bf q} d{\bf p}, \label{w1} \\
P&=&(2\pi)^M\int W^2\ d{\bf q} d{\bf p}\label{w2}.
\end{eqnarray}
Clearly, $C$ measures the structure of $W$.
Note that if we define $\alpha_m= (q_m+ip_m)/\sqrt{2}$, then Eqs.~(\ref{w1}) and (\ref{w2}) can be used to recover Eq.~(2) in Ref.~\cite{lee} (up to a factor of $2^M$ due to convention difference).

The $\chi^{2}$ measure well-established in the literature is defined as (see, e.g., Eq. (23) in Ref.~\cite{gu}, Eqs.~(16)-(19) in Ref.~\cite{gongpra}),
\begin{eqnarray}
\chi^{2} \equiv  \frac{2C}{P}.
\end{eqnarray}
Apparently, $\chi^{2}$ is determined by $C$ divided by purity $P$, whereas ${\cal I}$ is determined by $C$ minus a purity quantity $MP$.  For pure states, $P=1$, then ${\cal I}$ is equivalent to $\chi^{2}$, i.e.,
$
{\cal I}_{\text{pure}}= \chi^{2}/4- M/2.
$

For mixed states with unknown purity, ${\cal I}$ and $\chi^{2}$ do not have a one-to-one mapping.  It is hence necessary to discuss mixed states more.
Of particular interest are the mixed states examined in Ref.~\cite{lee} with ${\cal I}=0$, e.g., $\rho_{d}\equiv (1/d)\sum_{n=1}^{d}|n\rangle\langle n|$, where $|n\rangle$ are Fock states, or $\rho_{\alpha}=(|\alpha\rangle\langle\alpha|+|-\alpha\rangle\langle-\alpha|)/2$, where $|\pm\alpha\rangle$ are coherent states.   For convenience we assume $M=1$. In these cases Eq.~(\ref{i1}) gives $C=P$, from which we have $\chi^2=2C/P=2$.
At first glance it appears that the ${\cal I}$ measure is better because it yields zero for $\rho_{d}$ and $\rho_{\alpha}$ (``fully" mixed by intuition) whereas $\chi^2$ does not. However,
this feature is not really an advantage of ${\cal I}$. Specifically, $\chi^2$ is
positive-definite but ${\cal I}$ is not. As an example
consider a one-mode Gaussian Wigner function $W_{g}(x_1,p_1)=1/(\pi a^2)\exp[-(x_1^2+p_1^2)/a^2]$ with $a>1$.
For this mixed state one obtains $C=1/a^4$ and $P=1/a^2$.  Hence ${\cal I}=(1-a^2)/(2a^4) $ and $\chi^2=2/a^2$. It is seen that $0<\chi^2<2$ but ${\cal I}$ is now negative. Interestingly, either the ${\cal I}$ measure or the $\chi^2$ measure leads to the same conclusions: irrespective of their sizes (i.e., values of $d$, $\alpha$ or $a$),  states $\rho_{d}$ and $\rho_\alpha$ possess the same coherence, and state $W_{g}(x_1,p_1)$ should have even less coherence.

Parallel to $\chi^2$, a simple classical analog~\cite{gu,gongpra} of ${\cal I}$ can be defined. As a final note, phase space structure measure has also been extended to many-spin systems by use of bosonic representation of spin operators~\cite{vinitha}.


\begin{thebibliography}{99}
\bibitem{lee}C.~W.~Lee and H. Jeong, \prl{106}, 220401 (2011).
\bibitem{gu}Y. Gu, Phys. Lett. A {\bf 149}, 95 (1990).
\bibitem{brumer} A.~K.~Pattanayak and P.~Brumer, \pre{\bf 56}, 5174 (1997).
\bibitem{gongpra}J.~B.~Gong and P. Brumer, \pra{68}, 062103 (2003).
\bibitem{casati} V.~V. Sokolov, O.~V. Zhirov, G.~Benenti, and G.~Casati, Phys.
Rev. E {\bf 78}, 046212 (2008).
\bibitem{vinitha}V. Balachandran, G.~Benenti, G.~Casati, and J.~B. Gong, \pre{\bf 82}, 046216 (2010).
\end{thebibliography}
\end{document}